\documentclass[preprint,12pt]{{elsarticle}}

\usepackage{amsmath}
\usepackage{amsfonts}
\usepackage{amssymb}
\usepackage{makeidx}
\usepackage{graphicx}

\newcommand{\bit}{\begin{itemize}}
\newcommand{\eit}{\end{itemize}}

\def\benu{\begin{enumerate}}
\def\eenu{\end{enumerate}}

\def\btab{\begin{tabbing}}
\def\etab{\end{tabbing}}

\def\bit{\begin{itemize}}
\def\eit{\end{itemize}}
\def\beq{\begin{equation}}
\def\eeq{\end{equation}}
\def\bec{\begin{center}}
\def\eec{\end{center}}
\def\btable{\begin{tabular}}
\def\etable{\end{tabular}}
\def\beqr{\begin{eqnarray}}
\def\eeqr{\end{eqnarray}}

\def\om{\omega}

\def\eps{\epsilon}

\def\Dl{\Delta}
\def\sg{\sigma}

\def\btab{\begin{tabbing}}
\def\etab{\end{tabbing}}
\def\beqrs{\begin{eqnarray*}}
\def\eeqrs{\end{eqnarray*}}

\begin{document} 
\begin{frontmatter}

\title{A two-step method for retrieving the longitudinal profile of an electron bunch from its coherent
radiation}
\author[mu]{Daniele Pelliccia}
\author[fl]{Tanaji Sen\corref{cor}}
\address[mu]{School of Physics, Monash University, Victoria 3800, Australia}
\address[fl]{Accelerator Physics Center, Fermi National Accelerator Laboratory, Batavia, IL 60510, USA}
\cortext[cor]{Corresponding author}

\begin{abstract}
The coherent radiation emitted by an electron bunch provides a 
diagnostic signal that can be used to estimate its longitudinal distribution. 
Commonly only the amplitude of the intensity spectrum can be measured and 
the associated phase must be calculated to obtain the bunch 
profile. Very recently an iterative 
method was proposed to retrieve this phase. However ambiguities 
associated with non-uniqueness of the solution are always present in the phase 
retrieval procedure. Here we present a method to overcome the ambiguity problem
by first performing multiple independent runs of the phase retrieval procedure and 
then second, sorting the good solutions by mean of cross-correlation analysis. 
Results obtained with simulated bunches of various shapes and experimental 
measured spectra are presented, discussed and compared
with the established Kramers-Kronig method. It is shown that even when
the effect of the ambiguities is strong, as is the case for a double peak in 
the profile, the cross-correlation post-processing is able to filter out 
unwanted solutions. We show that, unlike the Kramers-Kronig method, the 
combined approach presented
is able to faithfully reconstruct complicated bunch profiles.
\end{abstract}


\end{frontmatter}

\section{Introduction}
Short electron bunches are generated at  photoinjectors such as the A0 photoinjector
at Fermilab \cite{Carneiro, Lumpkin}, at X-ray Free Electron Lasers such as the Linac Coherent Light Source (LCLS) at SLAC 
\cite{Huang} and 
will be a key requirement for high luminosity in the proposed International Linear Collider (ILC) \cite{ILC}. 
At the new ASTA photoinjector at Fermilab, 
now being commissioned, bunches shorter than a picosecond will be created using
a two stage compression scheme \cite{ASTA}. Streak cameras are used to measure 
bunch lengths down to the picosecond scale but shorter bunches require either
electro-optical methods \cite{Wilke}, transversely deflecting rf cavities \cite{Akre}
or methods which use the coherent radiation at infrared wavelengths emitted by the 
bunch. Such radiation can have different sources such as
synchrotron radiation, transition radiation, and  diffraction 
radiation to name a few. The radiation is coherent when the wavelength is 
comparable or longer than the bunch length. 
\newline
At the A0 photoinjector the coherent 
transition radiation (CTR) emitted when the electron beam traverses a thin 
metallic foil has been used to determine the longitudinal distribution of the electron beam. The distribution is reconstructed from the measured amplitude and
calculated phase of the intensity spectrum.
The phase calculation uses the Kramers-Kronig (KK) method which
 relates the real and imaginary parts of the spectrum via a Hilbert transform. While the KK method is
relatively straightforward and has been used extensively, see e.g. \cite{lai94, lai94b, lai95, lai97, schn97, miha06}, it has a few drawbacks which will be discussed below. It is therefore useful to develop an
alternative technique to reconstruct
the longitudinal distributions of short electron bunches. 

Very recently a different approach was proposed \cite{bajl13}. It is based 
on the \textit{iterative} retrieval of the phase from the knowledge of the measured 
amplitude and imposing suitable constraints which must be satisfied by the 
longitudinal bunch profile.  This approach belongs to a much wider class of iterative
methods that aim to indirectly measure a signal from the magnitude of its Fourier 
transform \cite{gerc72, fien78, fien82}. By retrieving the missing phase in Fourier 
space, the signal can be reconstructed unambiguously. In \cite{bajl13} such a 
procedure was employed to retrieve the longitudinal profile of simulated and measured
CTR spectra and a heuristic approach to optimize the iterative reconstruction was 
given. We note that a phase retrieval method was also used in reconstructing the
two-dimensional transverse density of an electron bunch from its coherent
transition radiation \cite{Marinelli}.
\newline
A fundamental issue with these iterative algorithms is that the underlying 
mathematical problem is inherently ill-posed \cite{paga06} meaning that a number 2N 
of unknown quantities (for instance amplitude and phase of the target function) must 
be reconstructed starting from only N measured values (i.e. the amplitude of the 
spectrum). In practice though, under reasonable assumptions on the object function to
be retrieved, the problem can be uniquely solved in two or more dimensions 
\cite{bate82} while fundamental ambiguities are still present in the one-dimensional 
case \cite{fien78, paga06, bate82}. \newline
This aspect was briefly discussed in \cite{bajl13}, showing that in some cases of 
experimental significance, the solution is in fact unique.
Nevertheless a more general approach, able to deal with the ambiguities is desirable.
This is the subject of this paper, which aims to extend the results obtained in 
\cite{bajl13}, discussing in greater detail the problem of uniqueness of the solution. We shall show that this is extremely relevant when the longitudinal bunch 
distribution has more than a single peak. In this case the fundamental ambiguities 
inherent to the 1D problem can play a non-trivial role. \newline
We proposed an approach inspired from a similar procedure derived for x-ray optics 
\cite{pell12}. This method is based on the \textit{a posteriori} selection of the 
results of the reconstruction of different algorithm runs, starting from independent 
random estimates. The selection is made on the basis of the cross-correlation that 
each reconstruction displays with a reference one. \newline
The paper is organized as follows: in Sec. 2 the basis of the KK method is reviewed 
and the limitations associated with it are discussed. Sec. 3 deals with the problem 
of the ambiguities in iterative phase retrieval, while Sec. 4 is devoted to the 
detailed description of our method. Sec. 5 reports the results obtained with our 
method applied to simulated bunch profiles of characteristic shape: Gaussian, 
Lorentzian, combinations of the two and a profile expected in a bunch compressor. 
The effect of noise is also discussed here. 
The results related to measured bunch profiles at the A0 photoinjector are shown in 
Sec. 6 and discussed in Sec. 7. Conclusions are drawn in Sec. 8. 

\section{The Kramers-Kronig method and its limitations}

We briefly discuss the KK method here, more complete discussions can be found
in \cite{lai94b, lai97}. Let $s(z)$ denote the longitudinal
distribution of an electron bunch, then its Fourier transform (FT) is
\beq
S(\om) \equiv \rho(\om) e^{i \phi(\om)} = 
\int s(z) \exp[i \frac{\om}{c}z] \; \mathrm{d}z.
\eeq
Here $\rho(\om)$ and $\phi(\om)$ are the amplitude and phase respectively of the 
complex function $S(\om)$. The bunch form factor $F(\om)$ is defined as 
\beq
F(\om) \equiv \rho^2(\om) = S(\om) S^*(\om).
\eeq
The form factor is related to the bunch intensity spectrum $I(\om)$ via
\beq
I(\om) = I_1(\om)[ N + N(N-1) F(\om)],
\eeq
where $I_1$ is the spectrum of a single particle and $N$ is the number of
particles in the bunch. In the limit of a large bunch population,
the intensity spectrum of coherent radiation $I(\om) \propto F(\om)$
or equivalently the amplitude $\rho(\om) \propto \sqrt{I(\om)}$.

Since $s(z)$ is a causal function, the real and imaginary parts of its FT
are related by the Hilbert transform. Using this relation, the phase can be written as the sum of two
terms $\phi(\om) = \phi_m(\om) + \phi_{B}(\om) $ where $\phi_m$ is known as the minimal phase and $\phi_{B}$ is known as the Blaschke phase. They can be written as:
\beq
\phi_m(\om) = -\frac{2\om}{\pi}
{\cal P}\int_0^{\infty}\mathrm{d}x \frac{\ln(\rho(x))}{x^2 - \om^2}, \;\;\;
  \phi_B(\om) = \sum_j \arg(\frac{\om - \om_j}{\om - \om_j^*}).
\eeq
In the expression for $\phi_m$, ${\cal P}$ denotes the Cauchy principal value while in the formula for $\phi_B$,   $\om_j$ is the $j$th zero of $S(\om)$
in the upper half of the complex plane. Only the zeros nearby to the real axis
have a meaningful impact on the phase. When the zeros are far away, 
the Blaschke phase is linear in the frequency but a
linear frequency shift corresponds to a spatial translation of the profile
and does not affect the shape of the bunch. If there are no nearby zeros,
then the minimal phase is a useful approximation to the total phase. It can
be written in a form without the apparent pole-like singularity 
\beq
\phi_m(\om) = -\frac{2\om}{\pi}
\int_0^{\infty} \mathrm{d}x \frac{\ln(\rho(x)/\rho(\om))}{x^2 - \om^2}.
\label{minphase}
\eeq
The bunch profile can be obtained from the minimal phase by an inverse
cosine transform,
\beq
s(z) = \frac{1}{\pi c}\int_0^{\infty}\rho(\om) \cos[\phi_m(\om)-\frac{\om}{c}z] \mathrm{d}\om.
\eeq
For a Gaussian bunch, the minimal phase is linear in $\om$ while for 
a general asymmetric bunch profile, $\phi_m$ is nonlinear in $\om$. 
\newline
The KK method suffers from two main limitations: (i) the modulus of the spectrum
 must be known in principle for all frequencies and (ii) the minimal phase must be
a good approximation to the total phase \cite{lai97}. In 
practice one needs to extrapolate  the measured spectrum to very low 
frequencies \cite{lai97} and calculate the phase  from 
Eq. (\ref{minphase}). \newline
The assumption of no nearby zeros is valid for Gaussian functions and
combinations of Gaussians but it is not valid for other shapes with long tails such 
as Lorentzians. For these shapes the Blaschke phase cannot be ignored and 
approximating the phase by the 
minimal phase does not result in an accurate reconstruction, as will be seen later. 
In a photoinjector with a bunch compressor, the longitudinal profile is
far from Gaussian and it not clear \textit{a priori} that the minimal phase approximation will be accurate. \newline
From the description above, the need for an alternative method is clear. 
The iterative phase retrieval approach can offer such an alternative: 
it is valid even in the presence of nearby zeros of $S(\omega)$ and can be
more robust against missing data. Nonetheless the iterative method also suffers from 
limitations, which will be discussed in the next section.

\section{Ambiguities in 1D phase retrieval procedures}

Iterative phase retrieval methods have been applied to 
numerous 2D problems in optics and crystallography \cite{mill90}, electron microscopy \cite{hump12} and 
x-ray imaging \cite{nuge10}. They are well suited when the direct imaging of a sample is not possible due to 
the lack of a suitable objective lens. For example, aberrations of the objective may prevent imaging at 
the diffraction limit or the efficiency of the objective lens makes a direct imaging procedure impractical.  \newline
The iterative method, in its original form is based on the fast Fourier transform (FFT) routine to 
numerically propagate the wave front from the sample plane to the detector plane and \textit{vice versa}. 
The algorithm is constrained by the measured data in the Fourier space, and by any available \textit{a priori} information in real space. Typically one imposes a "support" constraint in real space, namely imposing that the solution must be zero outside a certain region. Support information is in many cases sufficient to retrieve the phase and the algorithm is very robust against noise and also missing information \cite{spence02}. 
In most applications, phase retrieval algorithms aim at solving some imaging problem and therefore are 
naturally implemented with 2D data. When a 1D problem is to be treated,  fundamental ambiguities in the 
solution do exist \cite{bate82, fien78}, which we will briefly describe in the following. 
\newline
Let $s(t)=s(z/c)$ denote the bunch distribution in the time domain. It is straightforward to prove that for any
 given real numbers $t_{0}$ and $\alpha$, the amplitudes of the FT of the functions $s(t)$, $s(t+t_{0})$, $\exp(i\alpha)s(t)$ and $s^{*}(-t)$ are the same (the
 symbol $^{*}$ denotes complex conjugation). In other words if the distribution is 
changed with any of the above transformations, the measured spectrum 
$\rho^{meas}(\om)$ will be unchanged. While a
constant shift and a constant phase are actually trivial ambiguities, the ambiguity 
between $s(t)$ and $s^{*}(-t)$ is more complicated to handle. In imaging
it represents the so called "twin-image" problem (see \cite{guiz12} for a recent 
review). This problem corresponds to the stagnation of the iterative algorithm which is unable to converge on either  $s(t)$ or $s^{*}(-t)$ and keeps bouncing between a combination of the two. \newline
Fienup \cite{fien78} reported iterative retrieval of 1D functions pointing out that depending on the 
constraints applied, the 1D iterative procedure can have a unique solution or multiple yet correlated solutions. 
The presence of correlated solutions and the poor convergence due to the presence of 
twins is very significant for the reconstruction of longitudinal bunch profiles, as 
the set of possible solutions becomes very large. For instance when the distribution 
is asymmetric and 
composed of multiple, partially overlapping peaks, the relative heights and widths of
the peaks can be incorrectly reconstructed due to the presence of twins of possible solutions. \newline
Recently, a method was proposed to improve the reconstruction of a 1D signal from a 
single 
measurement of the magnitude of its FT \cite{pell12}. The method works in two steps. The first step is 
an iterative phase retrieval algorithm employing known numerical strategies such as hybrid input-output or 
error reduction \cite{fien82}. This procedure is repeated many times with different independent random 
choices of the initial phase $\phi(\omega)$ and the corresponding  set of independent
reconstructions is stored. At this point the second step is taken, consisting in a post-selection of the 
reconstructions on the basis of their cross-correlations with a reference 
reconstruction. 
The selected profiles are then averaged to produce the final result. 
Therefore, while 
ambiguities and poor reconstructions can still affect each individual solution, the post-selection ensures that only the 
significant reconstructions are considered. In the next section we describe in detail both steps of the method.

\section{Description of the method}

In this section we describe the strategies to 
implement the iterative phase retrieval procedure optimized for the measurement of an electron bunch profile. 
As anticipated, the method is divided in two steps, the first consisting of many independent runs 
of a suitable iterative phase retrieval algorithm, followed by result post-selection. 

\subsection{Iterative phase retrieval algorithm}
The aim is to reconstruct 
the electron bunch longitudinal distribution $s(t)$ from the knowledge of the measured spectrum amplitude 
$\rho^{meas}(\omega)$. 
The flow chart of the algorithm is depicted in Fig. \ref{fig: scheme}. In practice we deal with discrete arrays of data, therefore in the following we 
shall use the discrete notation $s_{j}$ to denote the $j$-th element of the array 
$s$. Lower-case symbols will denote arrays in real space (time domain) while
capital letters will indicate arrays in Fourier space (frequency domain)
The iterations start by defining a random array  $\phi^{(0)}$ and assign it as phase 
to the measured amplitude, thus building the complex-valued spectrum at the zero-th 
iterate: $S^{(0)}=\rho ^{meas} \exp(i \phi^{(0)})$.  Then the following sequence is 
iterated:
\begin{enumerate}
\item Inverse FT the function $S^{(0)}$ to get the zero-th estimate of the bunch longitudinal distribution $s^{(0)}$;
\item Impose the time domain constraints, by applying any \textit{a priori} information on the distribution, to get an updated time domain function $\tilde{s}^{(0)}$;
\item Calculate its FT, to get a subsequent estimate in the frequency domain $\tilde{S}^{(1)}$ and calculate its phase $\phi^{(1)}=\arg \left( \tilde{S}^{(1)} \right) $;
\item Update the frequency domain iterate by imposing the frequency domain constraint (i.e. the measured data), $S^{(1)}=\rho^{meas} \exp(i \phi^{(1)})$;
\end{enumerate}
The procedure is iterated until convergence is attained, i.e. the solution does not appreciably change with further iterates. To monitor the quality of the convergence two different figures of merit (FOMs) are used. The first accounts for the variation of the time domain function between subsequent iterates
\begin{equation}
\Delta_{k}=\sum_{j}\frac{\vert s^{(k+1)}_{j} \vert -\vert s^{(k)}_{j} \vert}{\vert s^{(k)}_{j} \vert}
\label{delta}
\end{equation}
where $k$ is the iteration index. 

We assume that the convergence is reached whenever $\Delta_{k}$ becomes less than some threshold value. 
In addition, a second figure of merit is employed defined as
\begin{equation}
\epsilon_{k}=\sum_{j}\frac{\vert S^{(k)}_{j} \vert -\rho^{meas}_{j}}{\rho^{meas}_{j}}
\label{epsilon}
\end{equation}
$\epsilon_{k}$ is used to monitor how close  the $k^{\textrm{th}}$ estimate of the amplitude of the frequency domain function -- before applying the 
frequency domain constraint -- is to the measured data. By minimizing both figures of merit we ensure 
that the algorithm is both converging and the solution is close to the actual 
measured data. \newline  
The constraint in the time domain is imposed by using all types of \textit{a priori} information that are available for the problem under study. 
For the results reported here we used a geometrical constraint, that means imposing the bunch length cannot exceed a certain 
duration. Such duration -- which we call "support" -- is especially important to guarantee the convergence 
of the algorithm. If it is too small the algorithm fails to converge because it cannot find a suitable 
solution fitting within the support size. If the support is too large it is likely that two or more  ambiguous solutions are present at the same time, resulting in a poor convergence \cite{guiz12}. 
The support can be approximately estimated by the auto-correlation 
function associated with the measured spectrum \cite{fien82b} but care must be used if the longitudinal bunch distribution has multiple adjacent 
peaks or long tails, as we shall discuss in the next section.
 \newline   
An additional simplification (that may be included as support) comes from the nature of the function to be retrieved.
The longitudinal profile of the electron bunch has to be necessarily described by a real and positive function. This fact reduces the ambiguity of 
Hermitian conjugation to a simpler time reversal. \newline
To update the solution at each step using the time domain constraints, we used 
several cycles, each composed of a combination of hybrid input-output (HIO) 
followed by error reduction (ER) algorithms \cite{fien82}. The HIO is known to 
ensure a faster convergence, being less prone to stagnations, while the ER 
steps are beneficial in reducing the reconstruction errors. 
\newline
A last remark concerns the possibility of dynamically updating the 
support to closely follow the object reconstruction. This 
corresponds to the so-called "shrink-wrap" algorithm \cite{marc03} and it has 
been successfully adapted to the phase retrieval applied to CTR \cite{bajl13}
where the authors were able to reduce the effects of ambiguities using this 
feature. This is indeed very beneficial when the distribution is composed of a 
single pulse with short tails (such as a Gaussian) or multiple well separated 
pulses. In this case the procedure will reduce the support size to the 
approximate size of the pulse (or the multiple pulses) and the algorithm will 
converge easily to the right shape. As we will discuss in the following, 
problems arise when the profile has long tails (as in a Lorentzian profile) or 
it is composed of partially overlapping pulses. In this case the support constraint cannot be very strict
and the convergence of the iterative procedure is generally poorer.
Therefore in this study we did not adopt a shrink-wrap option but introduced 
a solution-sorting procedure as a second step.  

\subsection{Post-selection of the reconstruction results}

\begin{figure}[h]
\centering
\includegraphics[scale=0.8]{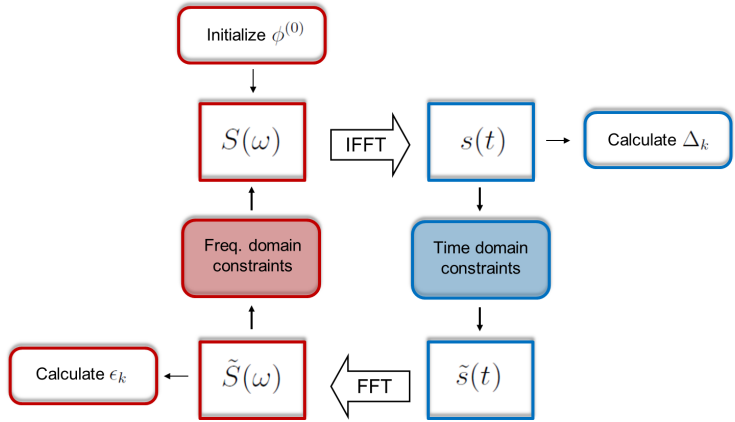}
\caption{Flow chart of the phase retrieval algorithm}
\label{fig: scheme}
\end{figure}

This procedure closely follows the one described in \cite{pell12} which we 
shall summarize here. From the first step of the method we obtain a set 
of $N$ independent results, each result being contained in an array with size $J$. 
Let us denote by 
$s_{j}^{n}$ the pixel $j$ of the $n$-th result, where $j=1,...,J$ and 
$n=1,...,N$. We shall denote with $s^{n}$ the array $n$. 
\newline
The aim of the post-selection procedure is to select the best results and 
average them, considering the figures of merit defined in the previous 
sub-section. Associated with each result are the values $\Delta$ and $\epsilon$ 
calculated with Eqs. (\ref{delta}) and (\ref{epsilon}) respectively at the last
step of each iteration run. The specific values of these FOMs vary depending on 
the profile to be reconstructed, typically values of 
$\Delta, \epsilon \lesssim 10^{-3}$ indicate
an acceptable convergence. If the values of the FOMs fail to meet this criterion the 
results are discarded, so that the set of $N$ results obtained 
at the end will meet the required level of convergence. 
\newline
The $N$ results will be affected by ambiguities described above, the most 
important in our case being translation shift, time reversal or a mild
problem related to the presence of twins.\newline
The key step at this point is to select the best solution obtained, seeking 
the lowest value of $ \epsilon$, indicating that the modulus of the reconstructed function in Fourier space is very close to the measured data. 
We then check that, associated with this solution, also that $\Delta$ has a low 
value, indicating that the algorithm did effectively converge. 
Typically when $ \epsilon$ is small, $\Delta$ is also small, while the converse is 
not always true, i.e. it is possible for the algorithm to converge to the wrong 
solution. 
The solution having the smallest value of $ \epsilon$ is chosen as the reference reconstruction, denoted by $s^{R}$ 
in the following. Every other solution is quantitatively compared to the 
reference by calculating the cross-correlation:
\begin{equation}
c^{n}_{h}=\frac{\sum_{j} s^{R}_{j}s^{n}_{j+h}}{\left( \sum_{j} \vert s^{R}_{j} \vert^{2} \right)^{1/2} \left( \sum_{j} \vert s^{n}_{j+h} \vert^{2} \right)^{1/2}}.
\label{cc1}
\end{equation}
The cross correlation is calculated for each solution as a function of the 
relative shift $h$. This procedure is necessary to account for any shift 
ambiguity that might be present. In practice $h$ is a cyclic shift, i.e. one 
always considers $(j+h) \mod J$. 
\newline
To account for the time reversal ambiguity a second cross correlation array is also computed:
\begin{equation}
\bar{c}^{n}_{l}=\frac{\sum_{j} s^{R}_{j}s^{n}_{J-j+l}}{\left( \sum_{j} \vert s^{R}_{j} \vert^{2} \right)^{1/2} \left( \sum_{j} \vert s^{n}_{J-j+l} \vert^{2} \right)^{1/2}}.
\label{cc2}
\end{equation}
In Eq. (\ref{cc2}) each array is reversed before computing the cross 
correlation.  In order to choose between ambiguous solutions 
one compares the cross correlations in Eqs. (\ref{cc1}) and (\ref{cc2}). In 
practice, denoting with $\bar{h}$ and $\bar{l}$ the values of the shift that 
maximize Eqs. (\ref{cc1}) and (\ref{cc2}) respectively, this corresponds to 
seeking the maximum between $c^{n}_{\bar{h}}$ and  $\bar{c}^{n}_{\bar{l}}$:
\begin{equation}
p_{n}=\max \left(c^{n}_{\bar{h}}, \bar{c}^{n}_{\bar{l}} \right).
\end{equation}
The value of $p_{n}$ indicates how well a solution 
is correlated with the reference. Therefore only solutions that after this 
procedure have a high correlation (say  $p_{n}\geq 0.9$) are averaged. 
Figure \ref{fig: scheme} shows the algorithmic flow of the method.

\section{Phase retrieval of simulated spectra}

\begin{figure}[h]
\centering
\includegraphics[scale=0.5]{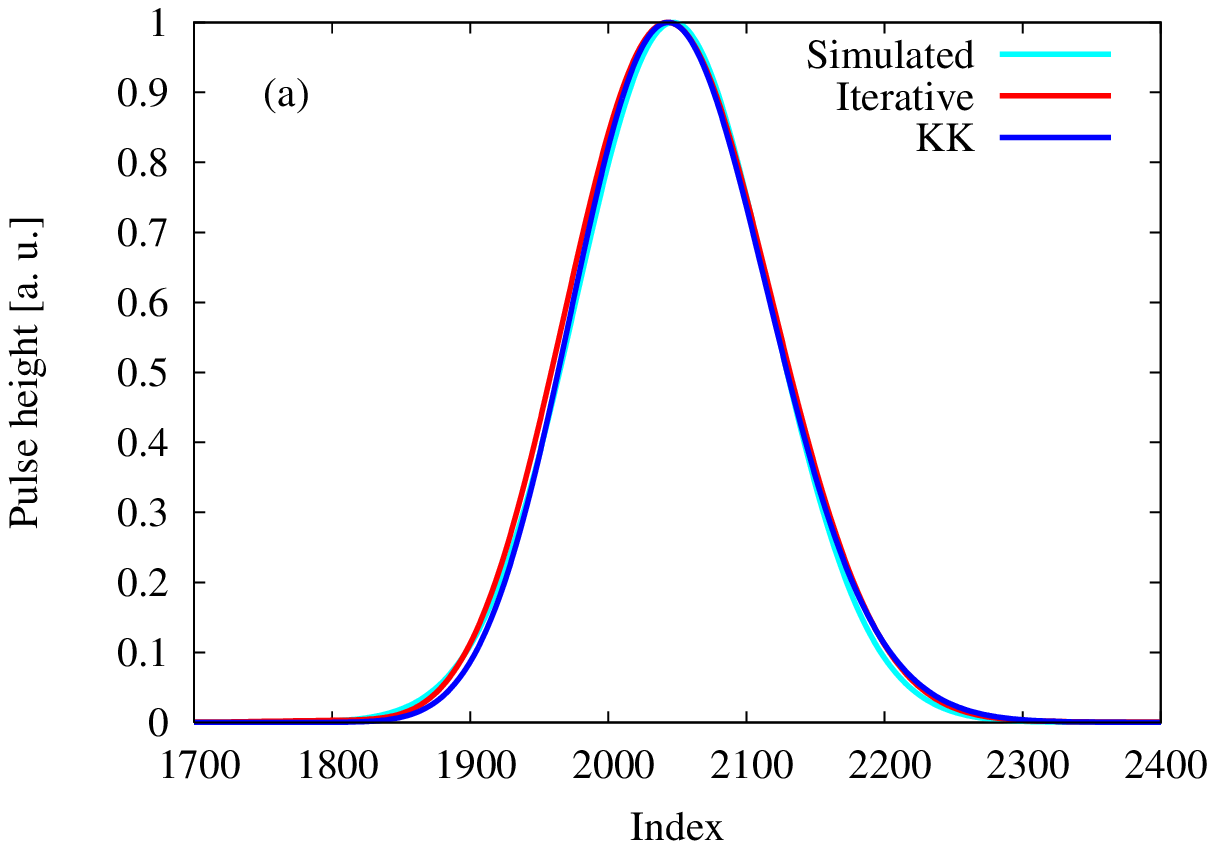}
\includegraphics[scale=0.5]{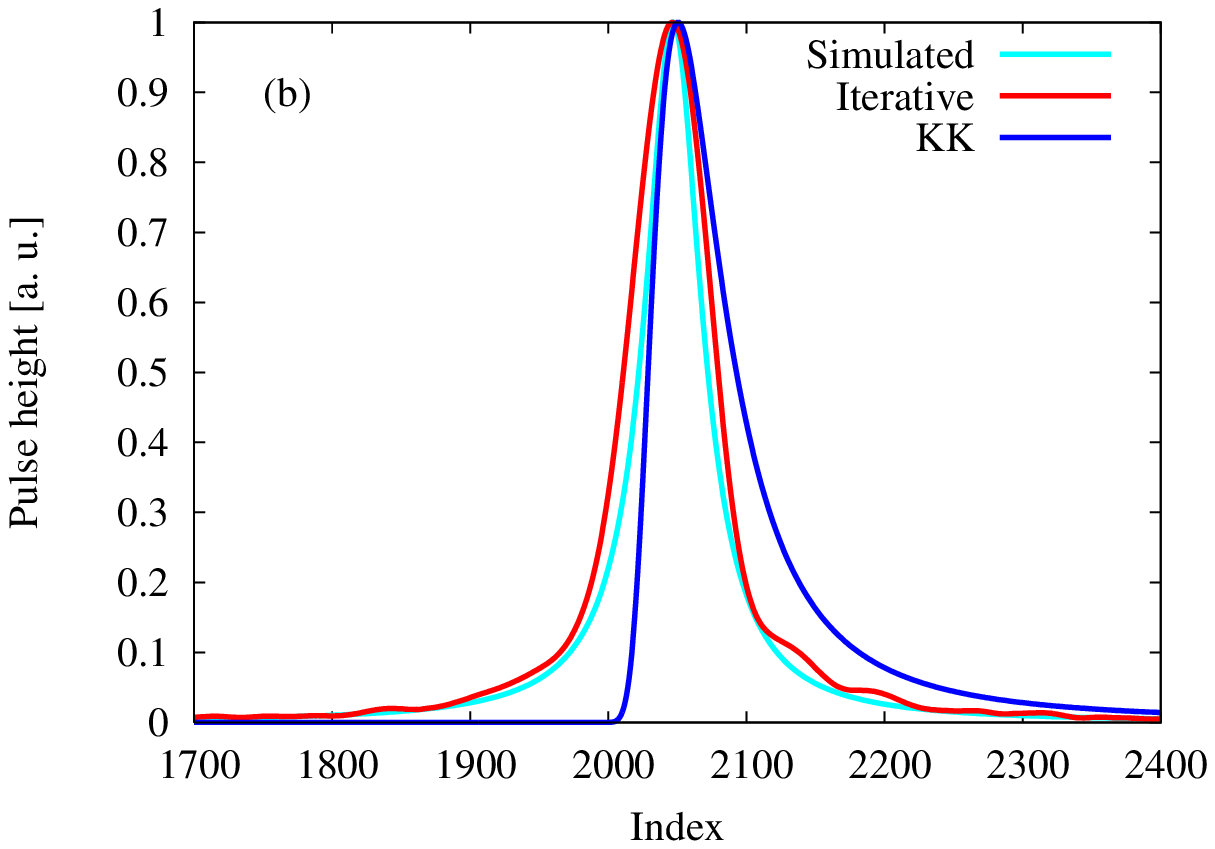}
\includegraphics[scale=0.5]{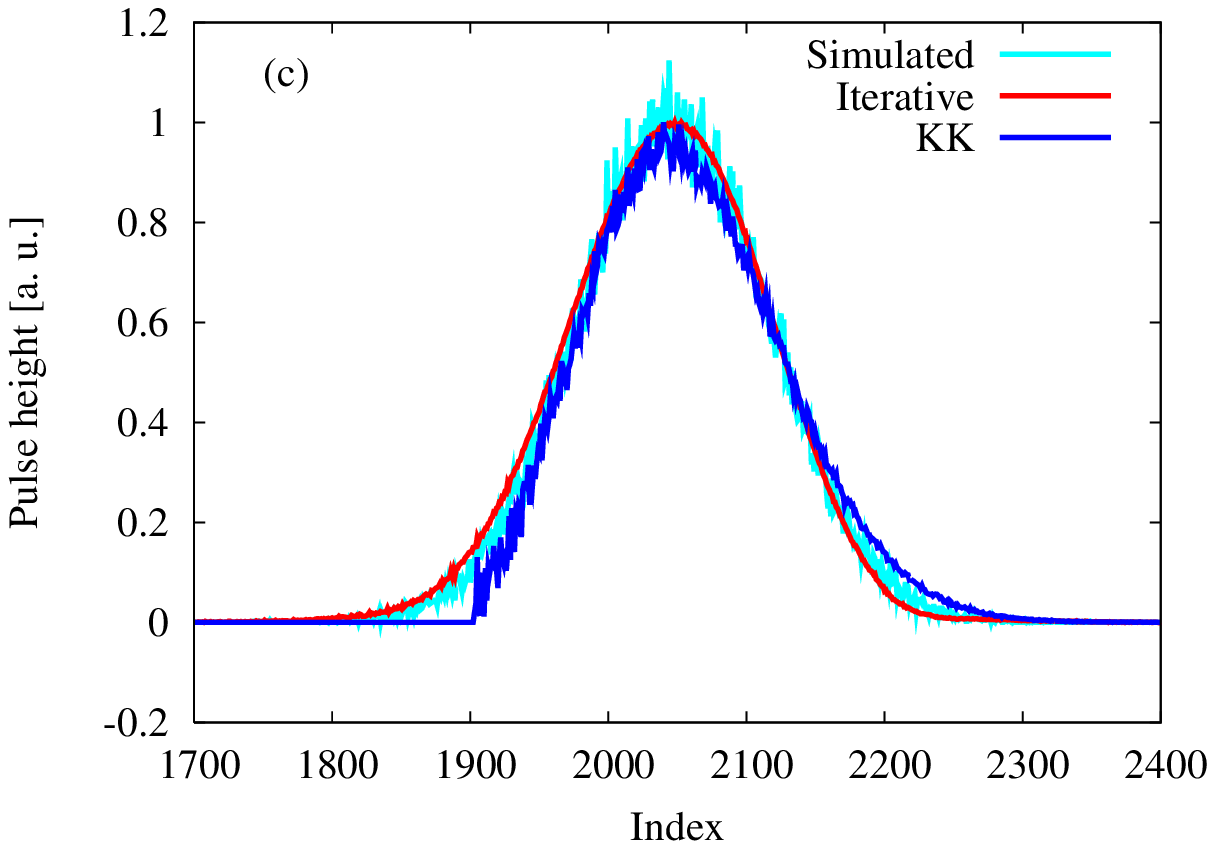}
\includegraphics[scale=0.5]{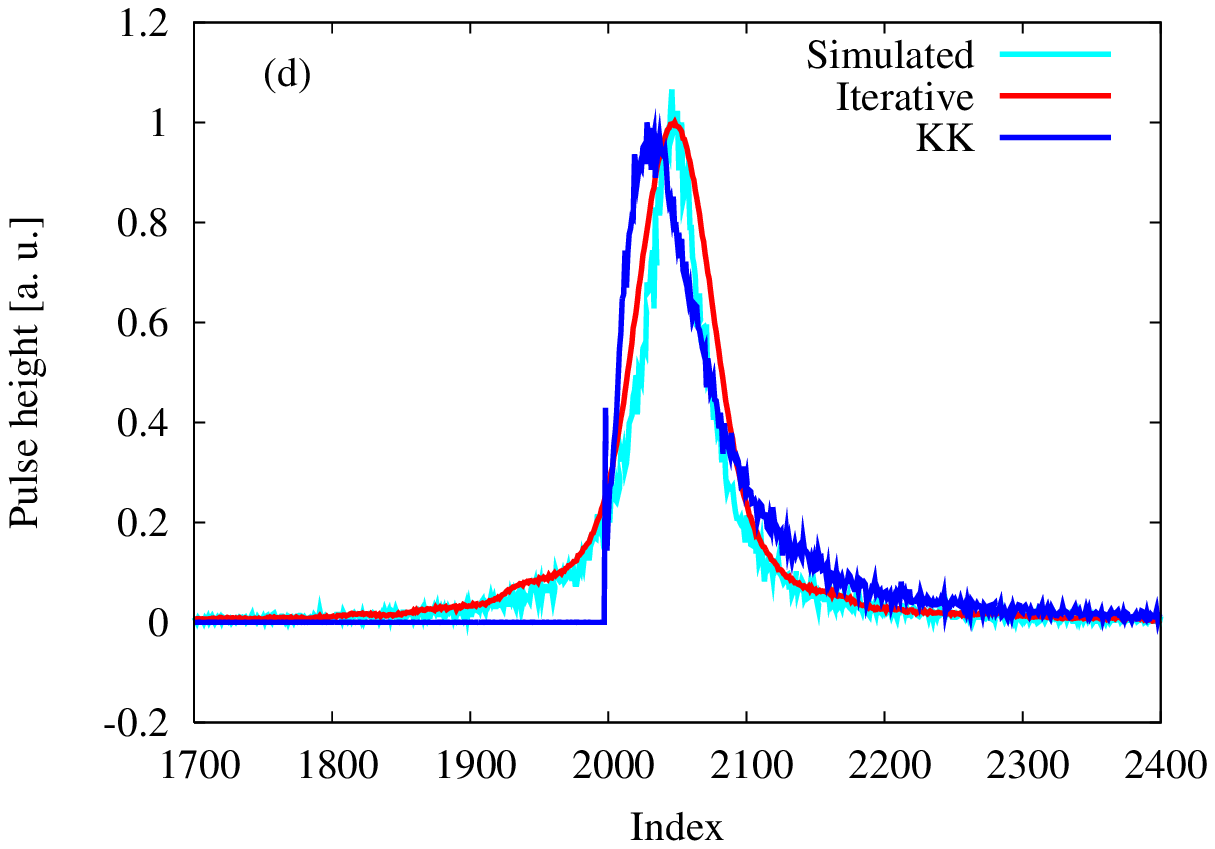}
\caption{Plots of the simulated profile (cyan), reconstructions using the KK 
method (blue) and the iterative phase retrieval (red). (a) Gaussian profile. 
(b) Lorentzian profile. (c) Gaussian profile with noise. (d) Lorentzian profile
with noise.}
\label{fig2}
\end{figure}

We performed the phase retrieval with the procedure described in the previous
 section using simulated spectra of four representative distributions: Gaussian,
Lorentzian, different superpositions of the two and a profile modeling that in
a bunch compressor. \newline
The Gaussian case is the easiest and very good reconstructions can be obtained 
both with the KK method \cite{lai95} and with the phase retrieval method
\cite{bajl13}, as shown in Fig. \ref{fig2}a. The Gaussian has been simulated using 
rms width $\sigma=$70 and its maximum amplitude value was set to unity.  \newline
On the contrary, the reconstruction of a Lorentzian pulse is non-trivial
in both scenarios, but for different reasons. 
The complex spectrum associated with a Lorentzian distribution has zeroes in the
upper half complex plane  close to the real axis, therefore the KK method 
(based on the minimal phase) is not able to reconstruct the pulse shape
properly. At the same time a Lorentzian curve has long tails, comparatively 
longer than a Gaussian. This fact makes the phase retrieval 
reconstruction based on geometrical support constraint also problematic, as 
the approximation of time-limited (bound) signal is not completely met. \newline
A satisfactory reconstruction is attained by increasing the size of the 
support 
(compared to the FWHM of the pulse). This improves the convergence even though some 
distortion is still present in the pulse tails, as evident from Fig. \ref{fig2}b. 
The FWHM of the simulated Lorentzian was $w=$50.   \newline
When random noise is added to the pulse, the performance of the iterative 
algorithm is still robust. Fig. \ref{fig2}c shows the Gaussian case with 
noise. 
It is worth noting that the KK reconstruction, which works well for a noiseless Gaussian, 
 has some issues when noise is added. This is not surprising considering that 
 the spectrum amplitude is no longer a smooth analytic function, so the
Hilbert transform relation between the amplitude and phase can only be approximately 
true.  
Using the iterative procedure, the retrieved profile with noise always appear 
smoother due to the averaging during the post-selection process. These observations 
are confirmed when a Lorentzian pulse with noise is simulated 
(Fig. \ref{fig2}d). The iterative algorithm is still capable of reproducing the tails
of the pulse correctly, while displaying a slightly larger width. 

\begin{figure}
\centering
\includegraphics[scale=0.5]{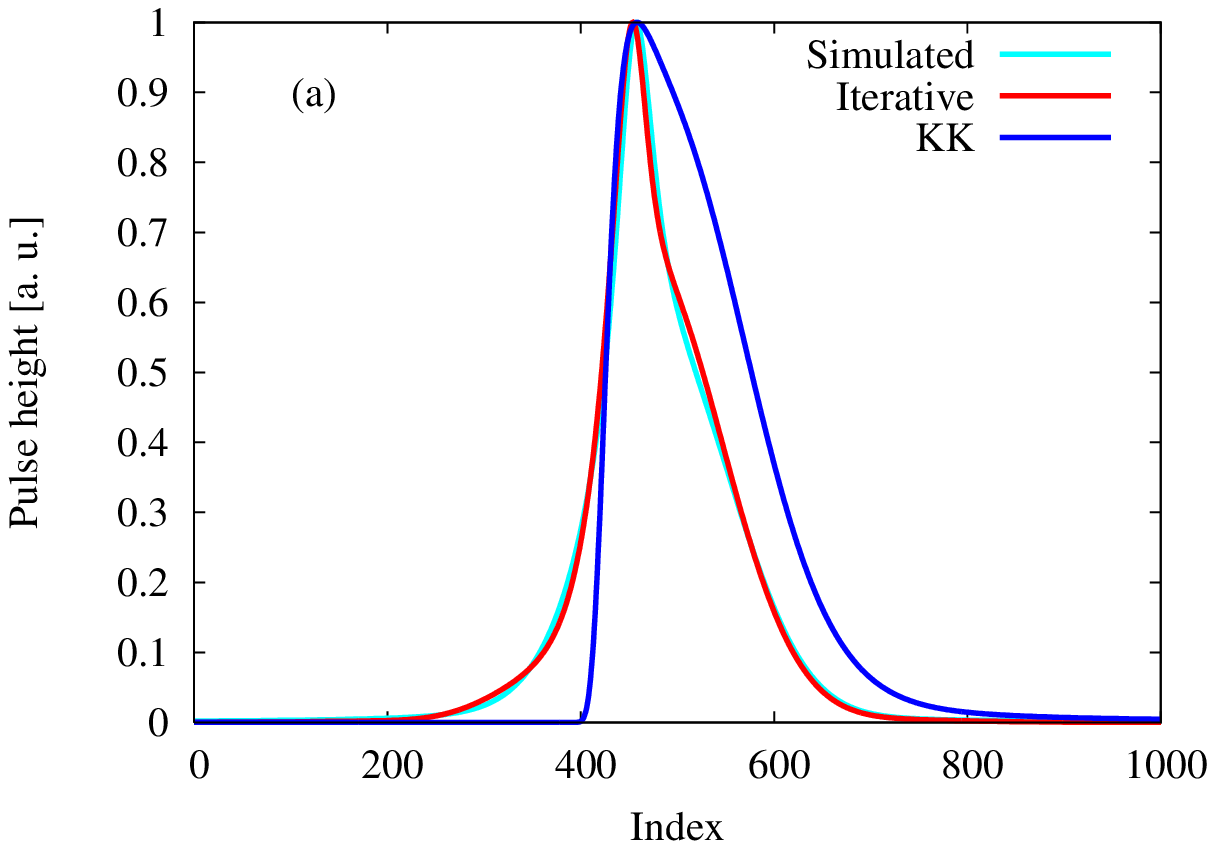}
\includegraphics[scale=0.5]{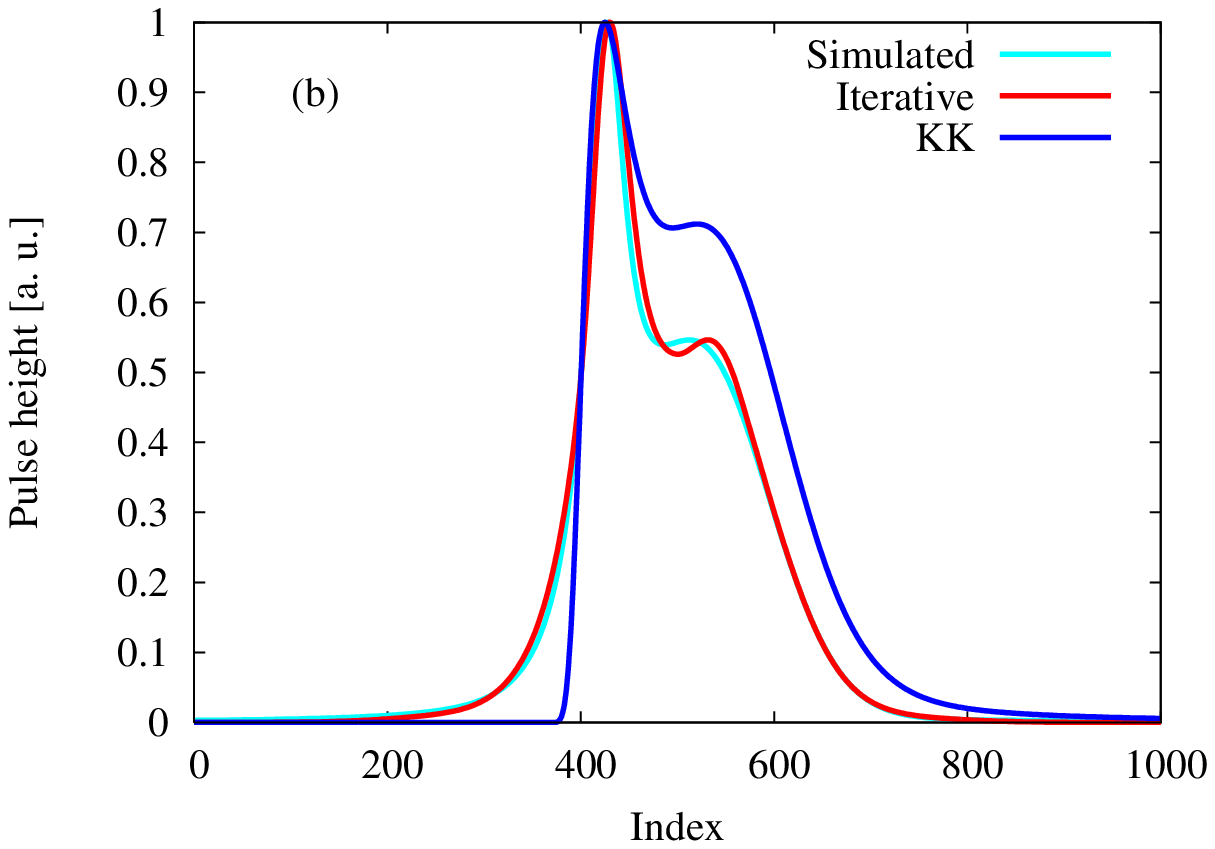}
\includegraphics[scale=0.5]{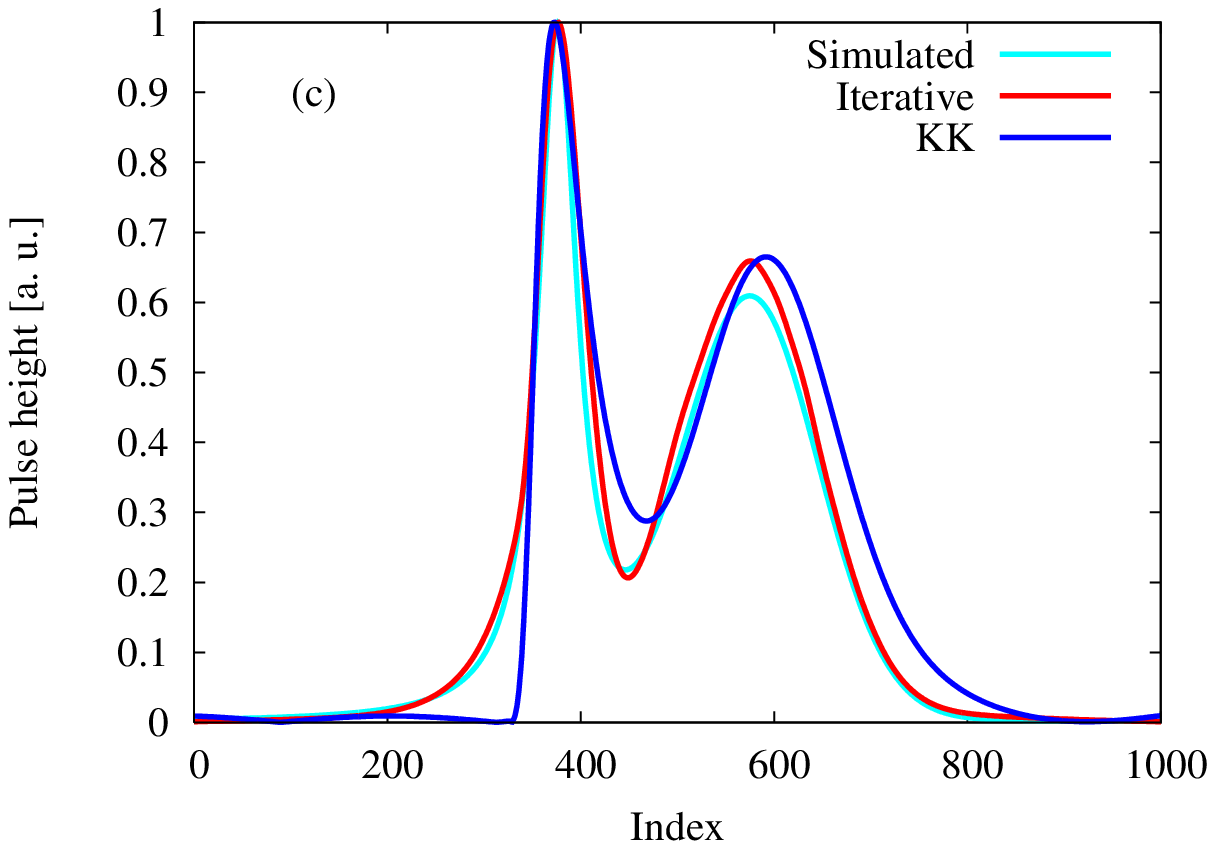}
\includegraphics[scale=0.5]{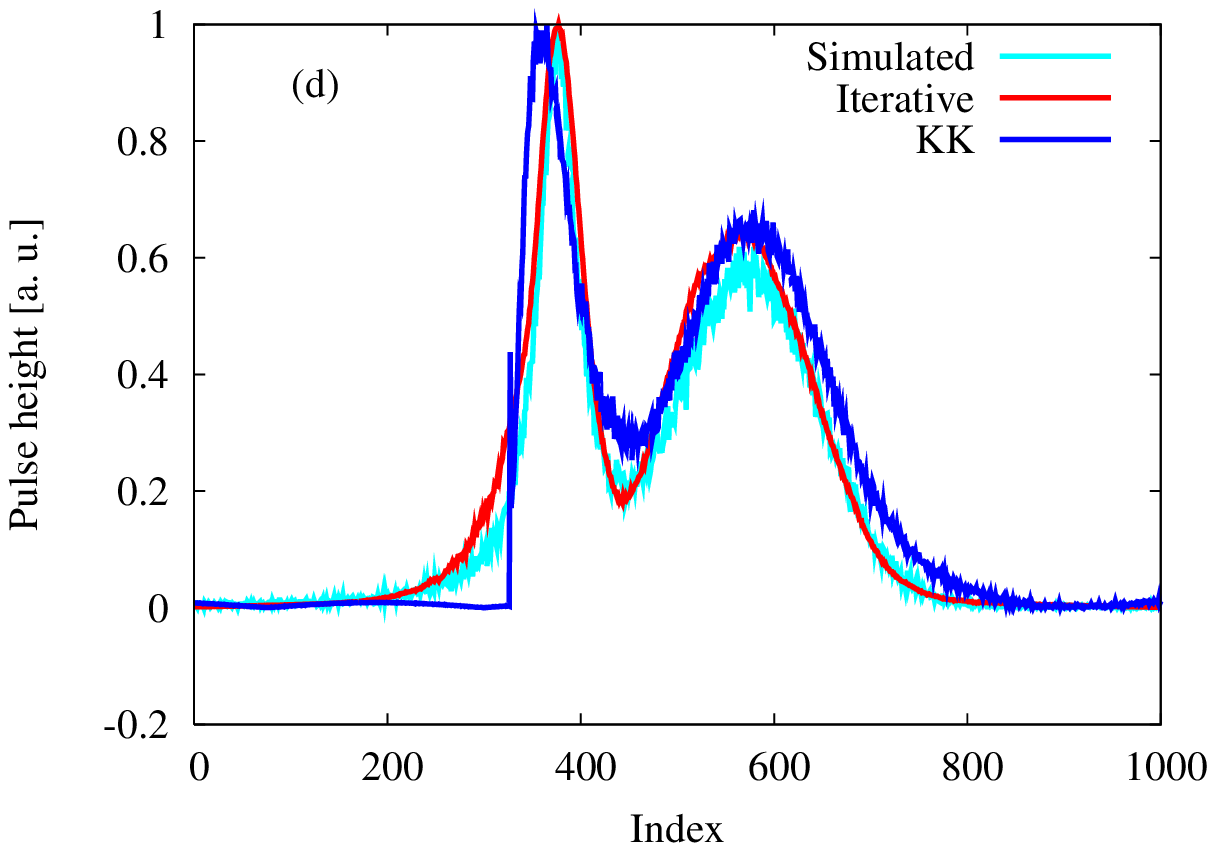}
\caption{Plots of the simulated profile (cyan), reconstructions using the KK 
method (blue) and the iterative phase retrieval (red). The profiles are the sum
of a Gaussian with $\sigma=70$ and a Lorentzian with width $w=50$ (see text). 
The distance $d$ between the peak centers is varied (a) $d=40$. (b) $d=100$. (c) $d=200$. (d) $d=200$ with noise added.}
\label{fig3}
\end{figure}

In Fig. \ref{fig3} the retrieved shapes from simulated spectra related to double 
pulses are shown. 
The simulated bunches contain different superpositions of a Gaussian and a Lorentzian shape with different degree of overlap, 
which can be typical of many experimental situations. 
In all cases we considered a Gaussian with $\sigma=$70
and a Lorentzian with width $w=$50. The maximum value of the Gaussian was  set
 to 60$\%$ of the corresponding value for the Lorentzian and the maximum height of the whole profile has been normalized to unity. We studied the effect of varying the separation
between the pulses and compared the results from the iterative reconstruction to those 
with the KK method. \newline
Fig. \ref{fig3}a shows the reconstruction obtained when the distance between the 
centers of the peaks is 40. In this case the profile appears rather asymmetric 
with different tails. The iterative reconstruction is very close to the
input profile, while an evident discrepancy appears for the KK reconstruction. A similar 
conclusion can be drawn from Fig. \ref{fig3}b where the separation between the peak 
centers is 100 but the two peaks are still partially overlapping. A value of 200 for 
the peak separation has been chosen for the results shown in Fig. \ref{fig3}c, without
noise, and in Fig. \ref{fig3}d, with noise, respectively. In this case the overlap 
between the pulses is much reduced and detrimental effects of time reversal ambiguities
appear to be stronger. In particular the relative heights of the peaks is not 
correctly reproduced and, when noise is present, distortions
in the tails do appear.

\begin{figure}
\centering
\includegraphics[scale=0.8]{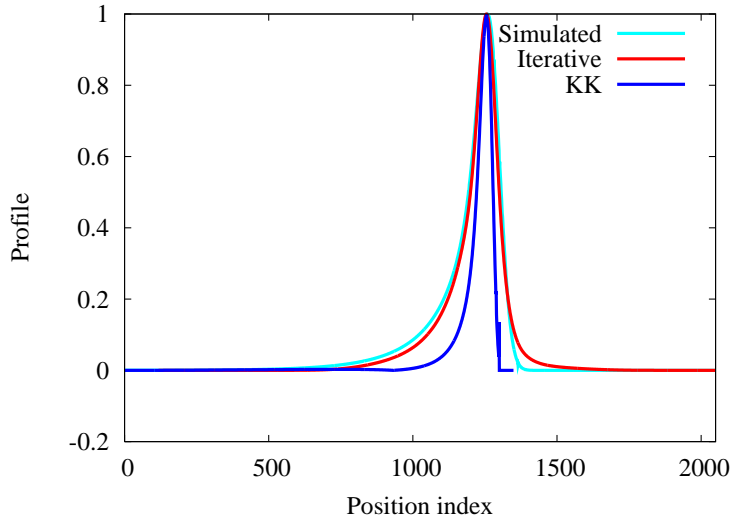}
\caption{Reconstruction of a bunch profile  expected (cyan) from a bunch 
compressor, see Eq.(\ref{eq: BC_prof}), with the iterative (red) and KK (blue) 
methods.}
\label{fig: BCprof}
\end{figure}

We have also tested the reconstruction with a theoretical model of the profile
expected in a bunch compressor \cite{Li}. The profile is given by the convolution
\beq
s[z,\sg_1,\sg_2] = \frac{1}{2\pi \sg_1\sg_2}
\int_{-\infty}^{\infty}dt \; \frac{\exp[(z-t)/(\sqrt{2}\sg_1)]}
{\sqrt{-(z-t)/(\sqrt{2}\sg_1)}} \Theta(-(z-t)) \exp[-\frac{t^2}{2\sg_2^2}]
\label{eq: BC_prof}
\eeq
where $\Theta$ is the Heaviside theta function and $\sg_1 > \sg_2$. The
input profile with $\sg_1=1.5, \sg_2=0.4$, the reconstructed iterative and
KK profiles are shown in Fig. \ref{fig: BCprof}.
As with the Lorentzian, the iterative reconstruction requires that the support be
large enough to accommodate the long tails. When that is done, the reconstructed
iterative profile here is very close to the input profile. By comparison, the KK
method yields a much narrower profile, falls off more quickly at both the head
and the tail and is not as accurate as the iterative profile. This last result
in particular shows that the iterative method is well suited for reconstructing
bunch profiles in a photoinjector.

\section{Experimental profiles at the A0 photoinjector}

The data used here were taken at Fermilab's A0 photoinjector during 2011. This was during a 
period when the Tevatron collider was
operational and liquid helium was available for a 9-cell superconducting cavity
in the beamline. 
We describe the experimental setup briefly here, more complete 
descriptions can be found in \cite{Frohlich, Thurman-Keup, Ruan, Thangaraj}. 
Electron pulses are generated with a CsTe photo-cathode within a 1.3 GHz rf gun
and then accelerated with a 1.3GHz superconducting rf cavity  to
an energy of about 15 MeV. Bunch charge could be varied over the range 
250pC - 1nC. A transverse to longitudinal emittance exchange (EEX) section 
consisting of a transverse 
deflecting mode cavity placed between two doglegs could be used to vary the 
bunch length by changing quadrupole settings upstream of the EEX \cite{Ruan}. 
In addition, the bunch length could also be controlled by
varying the off-crest rf phase (aka energy chirp) in the 9-cell
accelerating superconducting cavity which was upstream
of the EEX section. Downstream of the EEX, the beam passed through a thin 
metallic foil (Al coating on Si substrate) generating coherent transition 
radiation (CTR) in the process. The radiation entered
a Martin-Puplett interferometer which uses a polarizing splitter 
to send the radiation along two orthogonal arms with mirrors at the ends.
The mirrors change the polarizations and the beams after reflection are 
recombined using the same splitter
and then sent to two pyroelectric detectors after being split again 
with another polarizing splitter. One of the arms is movable so the
path length between the interfering beams can be varied. An interferogram
or auto-correlation function $C(t)$ is obtained from the intensity in the detectors; 
measured as a function of $t=\Dl z/c$ where $\Dl z$ is the path length 
difference from the central beam splitter to the two mirrors.
The real part of the Fourier transform of the
auto-correlation function yields the intensity spectrum $I(\omega)$.
The detector's response is frequency dependent and is limited at 
low frequencies by the thickness of the pyroelectric crystal. Consequently
the measured response drops to zero at frequencies below about 0.1 THz.
This is corrected during post-processing with a parabolic fit which
smoothly extends the response to zero frequency \cite{Thurman-Keup}. We use
the corrected intensity curves in the following.

\begin{figure}
\centering
\includegraphics[scale=0.5]{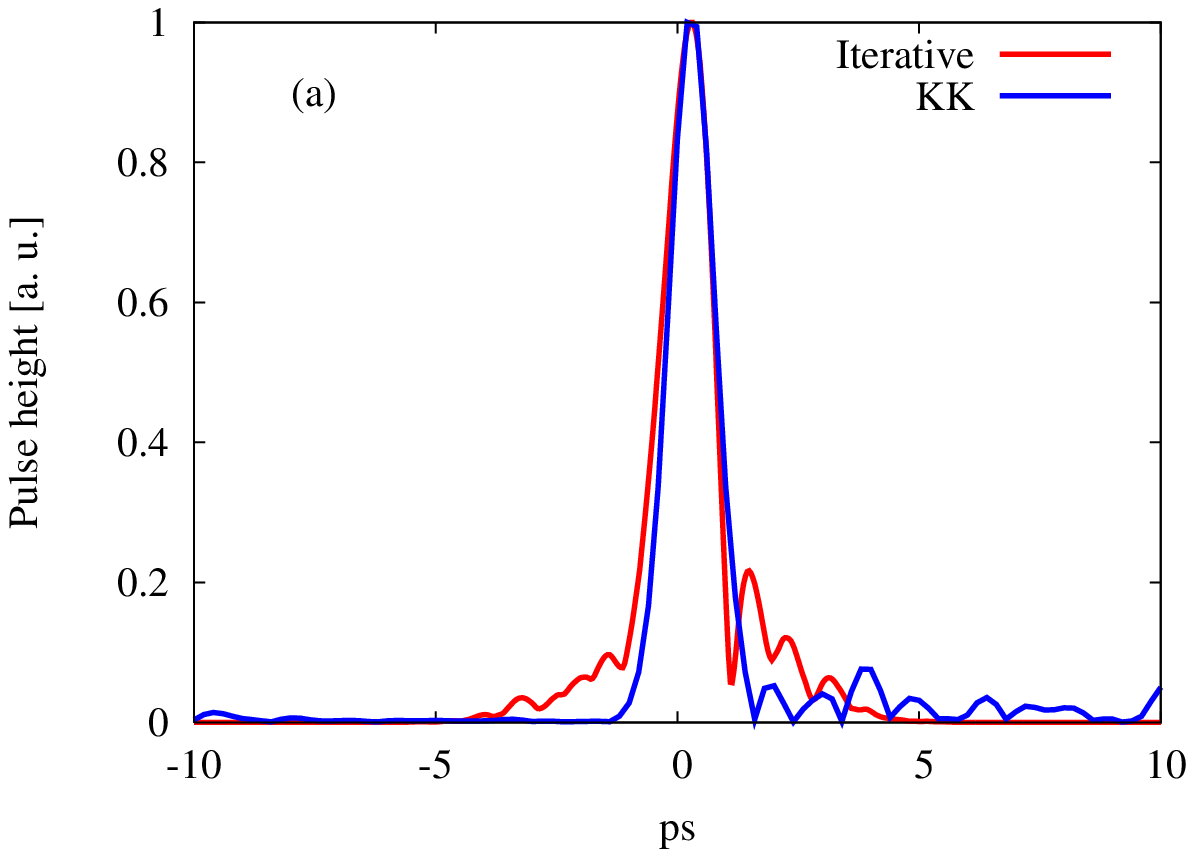}
\includegraphics[scale=0.5]{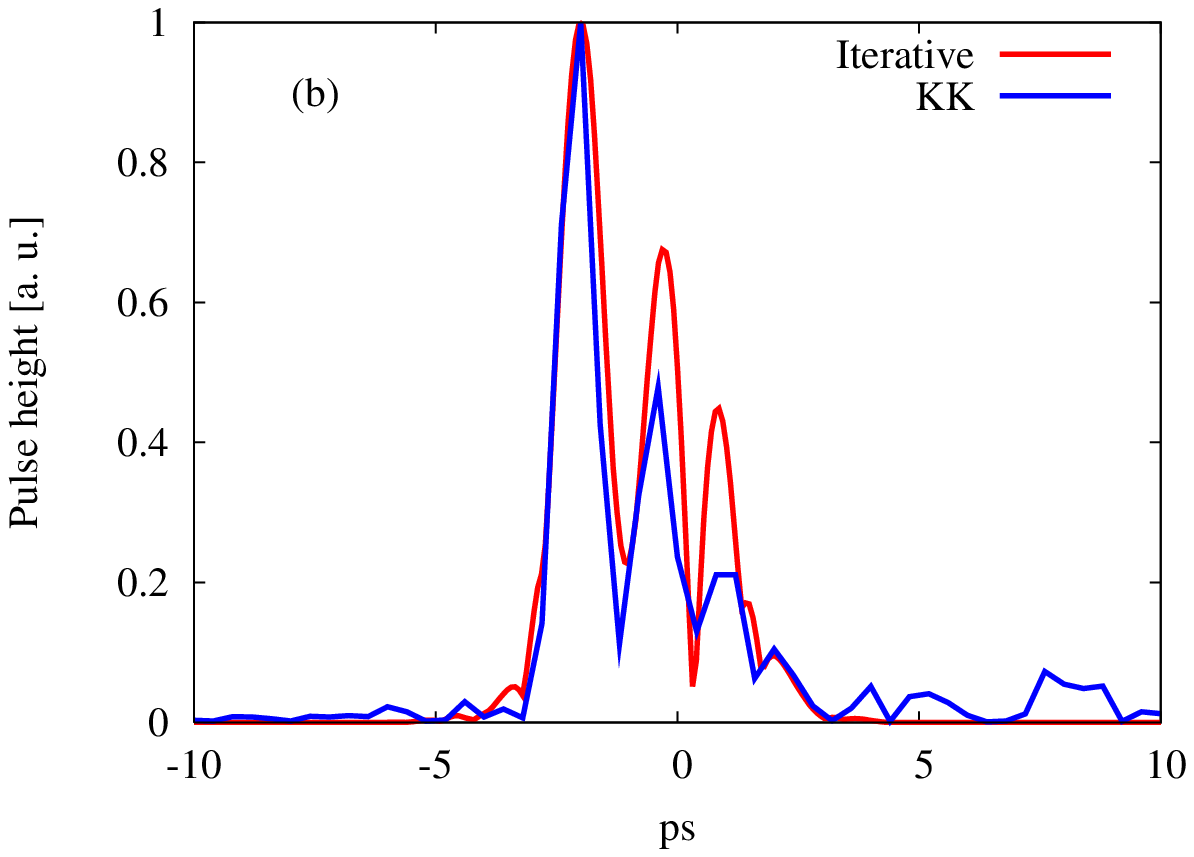}
\includegraphics[scale=0.5]{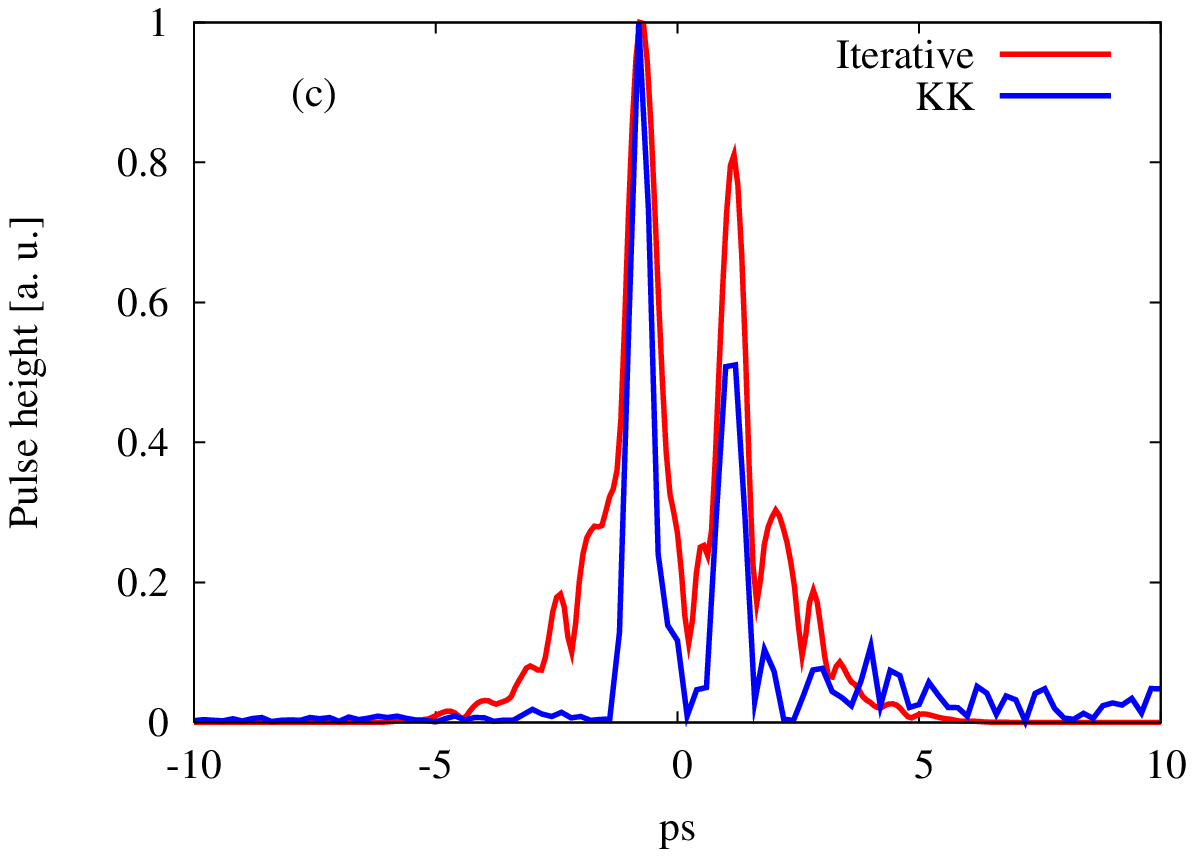}
\includegraphics[scale=0.5]{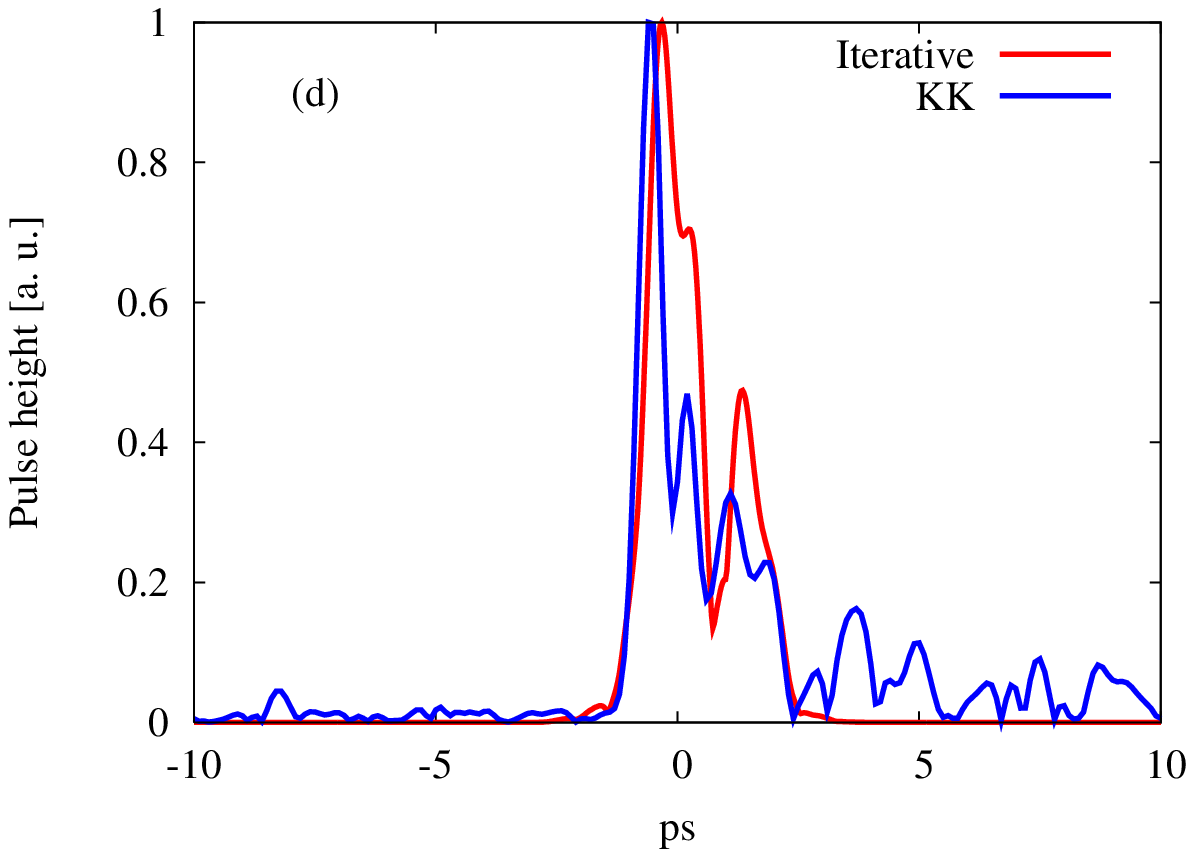}
\caption{Reconstruction of experimental data with iterative algorithm (red) 
and with the KK method (blue). Data were taken in 2011 and on different dates:
(a) May 9, (b) Aug 24, (c) Aug 25, (d) Oct 5 }
\label{fig5}
\end{figure}

Figure \ref{fig5} shows the reconstructed profiles from measurements taken over
four different days. In these measurements, the bunch length was varied only by
changing quadrupole settings upstream of the EEX and no energy chirp was 
applied. Both the KK and the iterative profiles are shown in Fig. \ref{fig5} and
they are scaled to the same peak height for ease of comparison. 

\begin{figure}
\centering
\includegraphics[scale=0.8]{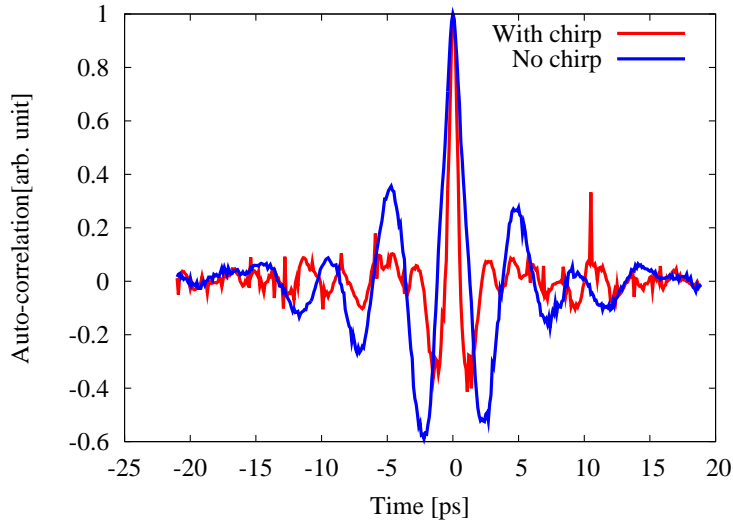}
\caption{Measured auto-correlation function of two pulses, data taken on
July 1, 2011. The data were taken with additional bunch compression using a
rf chirp (red) and also without the chirp (blue). See the text for a more detailed
description.}
\label{fig6}
\end{figure}

\begin{figure}
\centering
\includegraphics[scale=0.5]{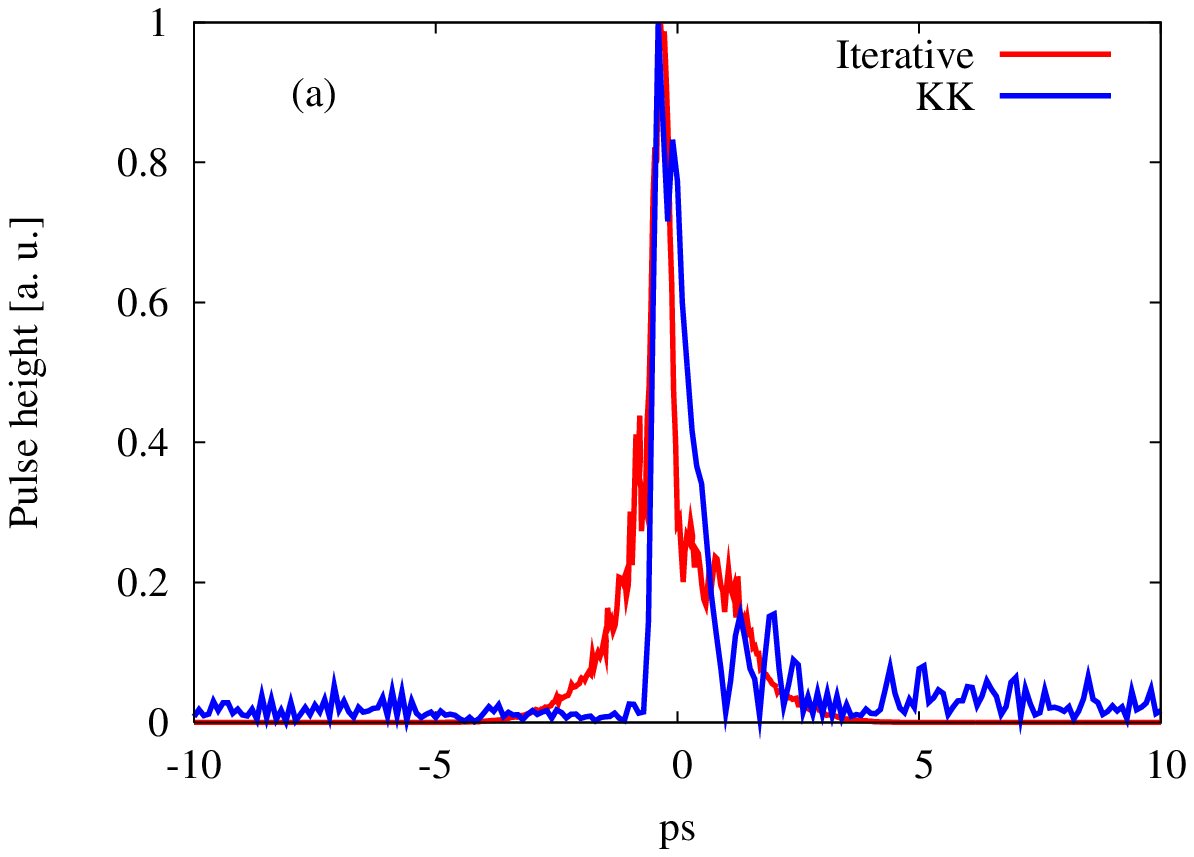}
\includegraphics[scale=0.5]{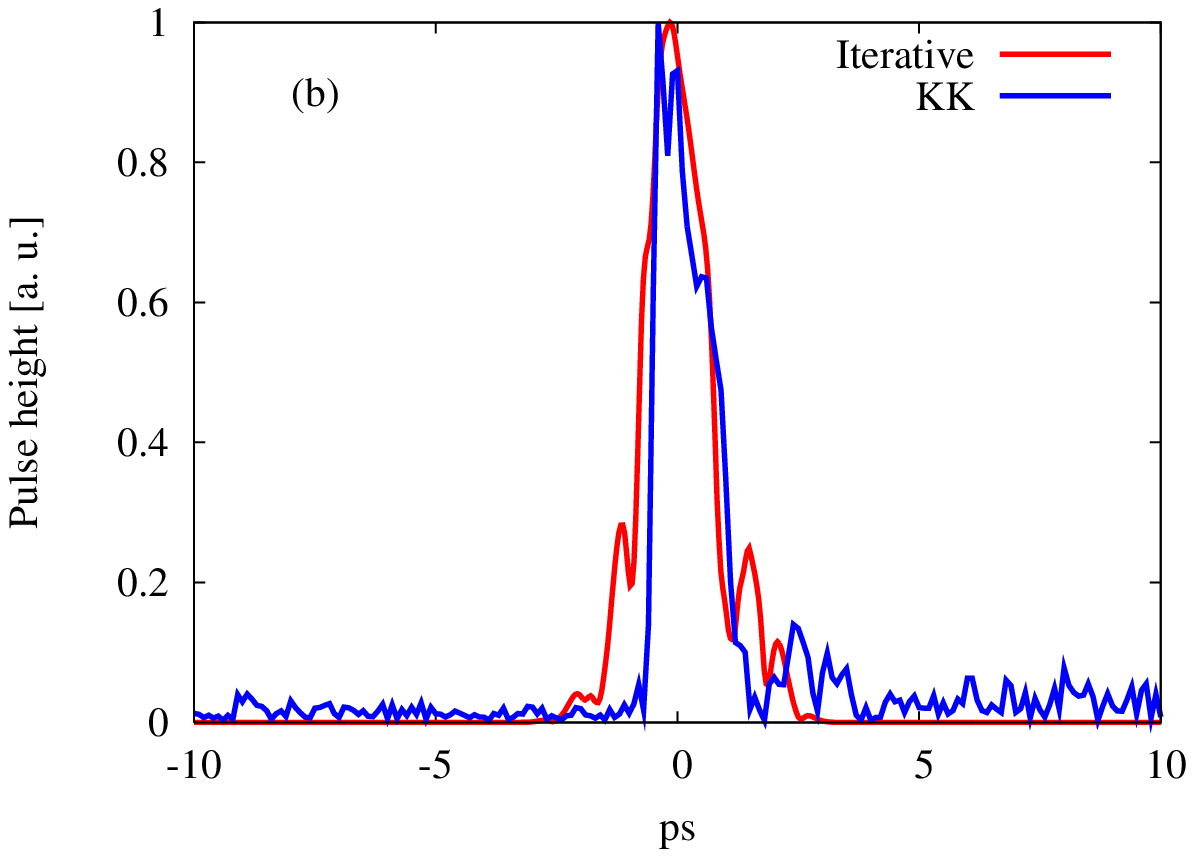}
\caption{Reconstruction of experimental data taken on July 1, 2011 with
the iterative method (red) and the KK method (blue)
(a) Compressed pulse with chirp, (b) Pulse without chirp.}
\label{fig7}
\end{figure}

A simple analysis shows that when the finite length of the rf cavity is
included, the longitudinal position of particles and therefore bunch profile
after the emittance exchange is
affected by the relative energy deviation prior to the exchange \cite{Thangaraj}. 
This was tested on another day when the effect of an energy chirp on the bunch 
length was measured by varying the off-crest rf phase. 
The coherent synchrotron radiation (CSR) emitted in the second dogleg was
monitored and at the maximum compression (17$^{\circ}$ off-crest) the CSR power
reached its peak value. A discussion of this measurement and the  CSR emitted 
by these bunches can be found in \cite{Thangaraj}. The auto-correlation 
function was measured with the chirp set at maximum compression and also
without any chirp and are shown in Fig. \ref{fig6}. As expected, the 
auto-correlation with the chirp has a significantly smaller full width at 
half maximum (FWHM). \newline
Figure \ref{fig7} shows the reconstructed profiles of 
these two cases, both with the iterative and the KK method. Again, the 
profiles have been scaled to the same peak height. 

\begin{figure}
\centering
\includegraphics[scale=0.8]{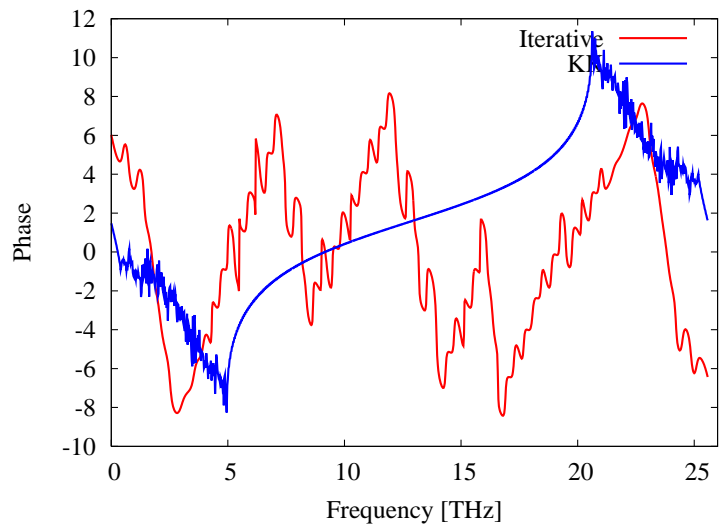}
\caption{Comparison of the minimal phase found with the KK method (blue)
and the nonlinear part of the phase from the iterative reconstruction for
the compressed profile with chirp.}
\label{fig: fig8}
\end{figure}

Figure \ref{fig: fig8} shows a comparison of the minimal phase used in the
KK method with the nonlinear (in frequency) part of the phase obtained with
the iterative reconstruction for the compressed profile. The difference may be 
attributed to the Blaschke contribution to the minimal phase. We note that 
both reconstructions will have the same autocorrelation function, since this
function is obtained from the Fourier transform of the intensity, which by
construction is the same for both methods. 

\section{Discussion of reconstruction results}
It is instructive to compare the quality of the reconstructions of simulated 
profiles to understand the merits and limitations of the iterative algorithm 
plus post-selection, over the KK method.
Both the  new method and the KK method perform very well for 
the reconstruction of the Gaussian profile alone -- Fig. \ref{fig2}a. 
The FWHMs from the two methods are shown in Table 1. \newline

The new method presented here tends to overestimate the FWHM while a slight 
under-estimate result from the KK. The reason for the over-estimate lies in 
the post-selection procedure. While averaging the best correlated solutions, a 
residual translation shift between them will result in an overall larger 
average. Nevertheless the effect is 
relatively small ($\sim 3\% $) and of little importance under normal 
circumstances.  \newline
The same effect is more significant in the reconstruction of the Lorentzian bunch, shown in Fig. \ref{fig2}b. The overestimation of the peak width is comparatively 
larger here because, in order to correctly reproduce the
long tails typical of the Lorentzian, the geometrical support constraint must 
be loosened, and this in turn produces a poorer localization of the peak. 
Therefore when averaging the results the presence of this residual shift 
ambiguity is more significant and yields a larger overestimation of the width. 
Nevertheless the profile shape closely follows the correct profile. The KK 
method on the other hand produces a steep vertical slope on one side and a
more slowly falling slope on the other side. These effects, arising from ignoring the
Blaschke phase, seems to be characteristic of the KK method when
reconstructing profiles with long tails, like Lorentzians.
\newline
From this discussion we conclude that the support size is an extremely critical
parameter, when reconstructing a bunch with long tails. We chose the optimal 
support size by seeking the best reconstruction in preliminary runs of the 
algorithm. 
If the support is too small compared to the profile width, the algorithm fails 
to converge and secondary small peaks may appear. Quantitatively this is 
manifested by both FOMs $\eps$ and $\Dl$ failing to decrease.
Therefore the procedure is to increase the support size, which automatically 
produces smaller value of $\Dl$ (i.e. increasing the support size makes the 
time domain constraint less and less effective). 
Therefore the optimal support size, on average, guarantees a low value of 
$\eps$ while avoiding stagnation of the algorithm. 
\begin{table}
\bec
\caption{ FWHM of the simulated profiles, compared with one reconstructed by KK
method and iterative retrieval.}
\btable{|c|c|c|c|} \hline
Profile type & FWHM input & FWHM KK & FWHM iterative \\ \hline
Gaussian & 165 & 163 & 170 \\
Lorentzian & 50 & 64 & 68 \\
Gaussian w noise & 165 & 160 & 170 \\
Lorentzian w noise & 50 & 68 & 64 \\
Bunch compressor profile & 114 & 57 & 100 \\
\hline
\etable
\eec
\label{table1}
\end{table}
The case of a double pulse is more convoluted. When two or more peaks are 
present, the time reversal ambiguity plays a much more critical role. 
Incorrect reconstructions often contain twins
resulting in a larger number of peaks or in an incorrect estimation of the
peak heights and widths. 
In this case the solution sorting procedure is a key factor
in filtering out incorrect reconstructions.   \newline
The comparison in Fig. \ref{fig3}, between the iterative reconstruction and 
the KK method shows that the former always produces a result much closer to the
input profile. Specifically, the peak positions are always correctly estimated.
\newline
Nonetheless the quality of the reconstruction critically depends on the peak 
separation. The iterative reconstructions are extremely close to the simulated
profiles for small peak separation (see Fig. \ref{fig3} a and b) while 
becoming worse when the separation increases. Notably the relative peak height 
is not faithfully reproduced. The dip between the peaks is 
more correctly reproduced with the iterative reconstruction. 
\newline
The same differences persist between the two methods in the reconstruction of
the theoretical profile in a bunch compressor given by Eq.(\ref{eq: BC_prof}).
The KK method produces a steeper vertical slope at the head and a FWHM half
that of the input profile. The iterative method profile matches the profile
closely both at the head and the tail and a FWHM close to that of the input.
\newline
It is worth noting that the cases studied here -- either the 
Lorentzian profile, the bunch compressor profile or the partially overlapping 
peaks -- are generally hard to
tackle with algorithms of the kind of ``shrink-wrap''. The double peak case is 
especially complicated as the shrink-wrap is able to restrict the support to 
the total extent of the profile, but ambiguities can still be present
in reconstructing the position/width of each individual pulse. 

\begin{table}
\bec
\caption{FWHM of the KK profiles compared with the FWHM
from iterative retrieval. In the case of bunches with multiple peak, the FWHM is relative the highest peak.}
\btable{|c|c|c|} \hline
Profile date & FWHM KK [ps] & FWHM iterative [ps]\\ \hline
May9 & 1.2 & 1.2 \\
Aug 24 & 0.8 & 1.1 \\
Aug 25 & 0.6 & 0.8 \\
Oct 5 & 0.6 & 1.2 \\
Jul 1, with chirp & 0.8 & 0.5 \\
Jul 1, no chirp & 1.4 & 1.5 \\
\hline
\etable
\eec
\label{table: fwhm}
\end{table}

The comparison of the experimental profiles in Fig. \ref{fig5} shows 
similarities and differences that are consistent over the measured profiles. 
Table \ref{table: fwhm} shows the FWHMs from the two methods for the
experimental profiles. 
Profile (a) from May 9 is 
seen to have only a single peak in both reconstructions. As before, the KK profile 
shows a very steep slope at the head of the pulse while the iterative profile
builds up more gradually. This feature is repeated in all the measured profiles
and was also seen in the simulated profiles of Figs. \ref{fig2}, \ref{fig3} and
\ref{fig: BCprof}.
The iterative profiles have higher secondary peaks in cases (b),(c),(d) in
Fig. \ref{fig5} and do not have the multiple small peaks in the tail 
when compared to the KK profiles. If real, these small peaks in the profile tails 
would indicate high frequency micro-structure 
These are unlikely to be present in the A0 photoinjector bunches given the 
relatively low charge and the lack of external excitations. It is therefore
possible that the iterative reconstruction avoids the appearance of
unphysical micro-structure, but this needs to be verified with detailed studies
in the ASTA photoinjector. \newline
 Comparing the profiles in Fig. \ref{fig7}, we find that both methods show
that the chirp reduced the pulse width relative to the unchirped case but the 
iterative profile is
narrower. The KK profile does not show that the additional compression with the
chirp reduced the FWHM compared to profiles obtained without chirp on other days, 
e.g. compared to the profiles seen in Figs \ref{fig5}c and \ref{fig5}d. The iterative
profile on the 
other hand shows that the chirp results in the smallest FWHM among all profiles.
This corresponds nicely to the auto-correlation function (seen in Fig \ref{fig6}) 
of the chirped pulse which also had the the smallest FWHM amongst all 
auto-correlations considered here. 
This is again suggestive, but not definitive evidence, that the iterative 
reconstruction may be more accurate. \newline
In concluding the section, it is worth commenting on the run time for the 
algorithm's execution. Clearly the number of steps required to perform the method
implies a larger run time when compared to the KK method. In our case the 
iterative algorithm was found to generally converge within 3000 iterations for 
the experimental data, with running time of about 635 ms on 1024 data points on a
single-CPU 3.3 GHz processor. The total time is then increased by the number of 
independent runs that are performed leading to a total run time of few minutes 
for 22 iterations including data I/O.
In comparison the KK method, which involves a simple FFT operation, completes
in about 1 ms on the same machine. Despite being longer, the run time of the 
iterative algorithm is still 
compatible with the use of the technique for on-line analysis of the longitudinal
properties of electron bunches.

\section{Conclusions}
 We have described a two-step iterative phase retrieval method to reconstruct the
phase of the spectrum from the measured spectrum amplitude. 
In the first step, multiple reconstructions are performed with
different initial random phases. During each reconstruction, constraints in 
frequency space and real space are applied successively. Only those solutions
are kept which are both sufficiently close to measured spectrum amplitude 
and have a high rate of convergence. In the second step, a post
selection method using cross-correlations is employed to remove ambiguities
related to translation and reversal in real space. The final solution is
obtained by averaging over the solutions surviving the post-selection step.
The post-selection is crucial in removing these troublesome ambiguities that
are common with complicated profiles such as those with multiple peaks.

In applying this procedure to Gaussians, Lorentzians, sums of Lorentzians and
Gaussians and a profile expected from a bunch compressor, we found that this 
iterative method is able to retrieve the
profiles with reasonable accuracy and much more faithfully than the 
Kramers-Kronig method based on the minimal phase. The latter method is known to
be inaccurate for profiles with long-tails. When applied
to experimental data taken at the A0 photoinjector, the iterative method
yielded profiles without some of the unphysical features in the profiles
of the minimal phase KK method. 

The major drawback of the iterative method at present is its iterative nature,
so it takes longer and requires some trial and error before the optimal 
solution is found. With some additional work, this procedure could be
automated and made available in control room applications.

\vspace{2em}

{\noindent \bf Acknowledgments} \newline
D.P. acknowledges the support of the Australian Research Council.
T.S thanks Charles Thangaraj and Randy Thurman-Keup for generously 
sharing their data and useful discussions. 
Fermilab is operated by Fermi Research Alliance, LLC under Contract 
No. DE-AC02-07CH11359 with the United States Department of Energy.


\end{document}